\documentclass{sigchi}

\CopyrightYear{2020}
\setcopyright{rightsretained}
\doi{TBD}
\isbn{TBD}
\conferenceinfo{UNDER REVIEW}{}
\acmPrice{}

\usepackage{balance}       
\usepackage{graphics}      
\usepackage[T1]{fontenc}   
\usepackage{txfonts}
\usepackage{mathptmx}
\usepackage[pdflang={en-US},pdftex]{hyperref}
\usepackage{color}
\usepackage{booktabs}
\usepackage{textcomp}
\usepackage{subfig}

\usepackage{microtype}        
\usepackage{ccicons}          

\usepackage{todonotes}

\def\plaintitle{Advertisers Jump on Coronavirus Bandwagon: \\Politics, News, and Business}

\def\emptyauthor{}

\makeatletter
\def\url@leostyle{%
  \@ifundefined{selectfont}{
    \def\UrlFont{\sf}
  }{
    \def\UrlFont{\small\bf\ttfamily}
  }}
\makeatother
\def\pprw{8.5in}
\def\pprh{11in}

\definecolor{linkColor}{RGB}{20,153,44}
\hypersetup{
  pdftitle={\plaintitle},
  pdfauthor={\emptyauthor},
  pdfkeywords={}, 
  pdfdisplaydoctitle=true, 
  bookmarksnumbered,
  pdfstartview={FitH},
  colorlinks,
  citecolor=black,
  filecolor=black,
  linkcolor=black,
  urlcolor=linkColor,
  breaklinks=true,
  hypertexnames=false,}

\newenvironment {squishlist}
{\begin{list}{$\bullet$}
  { \setlength{\itemsep}{0pt}
     \setlength{\parsep}{2pt}
     \setlength{\topsep}{3pt}
     \setlength{\partopsep}{0pt}
     \setlength{\leftmargin}{1.5em}
     \setlength{\labelwidth}{1em}
     \setlength{\labelsep}{0.5em} } }
{\end{list}}

\definecolor{kk}{RGB}{150, 0, 150}

\definecolor{ym}{RGB}{50, 200, 0}

\begin{document}

\title{\plaintitle}

\numberofauthors{1}
\author{%
  \alignauthor{Yelena Mejova and Kyriaki Kalimeri\\
    \affaddr{ISI Foundation, Turin, Italy}\\
    \email{yelenamejova,kkalimeri@acm.org}}\vspace{0.5cm}
}

\maketitle

\begin{abstract}

In the age of social media, disasters and epidemics usher not only a devastation and affliction in the physical world, but also prompt a deluge of information, opinions, prognoses and advice to billions of internet users. The coronavirus epidemic of 2019-2020, or COVID-19, is no exception, with the World Health Organization warning of a possible `infodemic' of fake news. In this study, we examine the alternative narratives around the coronavirus outbreak through advertisements promoted on Facebook, the largest social media platform in the US. Using the new Facebook Ads Library, we discover advertisers from public health and non-profit sectors, alongside those from news media, politics, and business, incorporating coronavirus into their messaging and agenda. We find the virus used in political attacks, donation solicitations, business promotion, stock market advice, and animal rights campaigning. Among these, we find several instances of possible misinformation, ranging from bioweapons conspiracy theories to unverifiable claims by politicians. As we make the dataset available to the community, we hope the advertising domain will become an important part of quality control for public health communication and public discourse in general.

\end{abstract}





\printccsdesc

\section{Introduction}

The coronavirus disease COVID-19 started in December 2019 in Wuhan, the capital of Hubei, China. At the time of the data collection for this paper on February 20, 2020, over 72 thousand cases have been recorded in China, including over 1,870 deaths, and around 700 people, mostly travellers, were diagnosed in the rest of the world \cite{who2020report}. Although the number of cases detected outside China remains much smaller than inside, the world's media and public attention remain focused on the ongoing developments. Prompted by a whirlwind of mainstream and social media coverage, the World Health Organization (WHO) has warned of a possible `infodemic' -- incorrect or malicious information being spread quickly and to a wide audience\footnote{\url{https://www.bbc.com/news/technology-51497800}}, while major social media platforms have pledged to use their networks third-party fact-checkers to ensure quality of information available to the public\footnote{\url{https://about.fb.com/news/2020/01/coronavirus/}}. Adapted to the agendas of various actors, the narrative around the epidemic is largely fragmented. Such fragmentation competes and sometimes contradicts, the fact-based educational messaging of the public health organizations.

Despite the best efforts of fact-checking organizations, the coronavirus outbreak has been taken up by social and political movements as an opportunity to communicate their messages. Outside China, xenophobic attacks have increased against businesses and individuals of Asian origin, prompting public statements from political leaders condemning racism\footnote{\url{https://www.economist.com/china/2020/02/17/the-coronavirus-spreads-racism-against-and-among-ethnic-\\chinese}}. Further, unscrupulous parties are taking to the social media platforms to promote dubious claims of remedies and cures\footnote{\url{https://www.scmp.com/week-asia/health-environment/article/3049261/garlic-cant-keep-coronavirus-bay-neither-will}}, while politicians are propagating conspiracy theories about the origins of the epidemic\footnote{\url{https://www.nytimes.com/2020/02/17/business/media/coronavirus-tom-cotton-china.html}}. Adapted to the agendas of various actors, the narrative around the epidemic is largely fragmented. Such fragmentation competes and sometimes contradicts, the fact-based educational messaging of the public health organizations.

In this work, we examine a novel dataset of advertisements posted to Facebook at the time of the epidemic's beginning in order to answer the following questions: 

\begin{squishlist}
\item How is Facebook advertisement used as a platform for conversation around coronavirus?
\item What alternative narratives are present around coronavirus outside public health messaging? Is this messaging emotionally provoking?
\item Finally, do the alternative narratives contain possible misinformation?
\end{squishlist}

The resource we use is the Facebook Ads Library, which was launched by Facebook in March 2019\footnote{\url{https://about.fb.com/news/2019/03/a-better-way-to-learn-about-ads/}} as a way for the public to ``learn more about ads related to politics or issues that have run on Facebook or Instagram''. Accessible through Facebook Ads Library API\footnote{\url{https://www.facebook.com/ads/library/api}}, it provides the title and text of the advertisement, the ID of the Facebook page and funding entity. Alongside these basic descriptors, it shows a demographic distribution of users reached by the ad in terms of age groups and genders, as well as a range of funds spent on the campaign. The data is available for the US, EU countries, and a handful of others, including Brazil, Israel, and Ukraine. 

We find a variety of advertisers invoking the epidemic, from public health and non-profit organizations updating their audience on the latest news and soliciting donations, to political and business entities coopting the threat of epidemic to their messaging and profit. Coinciding with the Democratic primary season in the US, the virus is often mentioned in association with political figures -- both in support and in opposition. Despite Facebook's near-ubiquitous reach across the US, these ads often target very specific demographics and locales, especially favoring California, New York and Texas. Furthermore, we find a range of possible erroneous information within these ads, ranging from conspiracy theories about bioweapons, to milder claims of political mismanagement and misunderstanding. However, as the situation develops, more will be known about the veracity of some of the content we discovered. Thus, we make the full dataset, along with the manual labels, available to the research community, in accordance with the Terms of Service of Facebook.

Findings of this study have wide implications for public health messaging. As we discover, there is a strong competition for the audience and the framing around the epidemic on the side of politics and news, potentially supplanting or contradicting the messages public health organizations may promote. Employing the epidemic as a tool for political attacks may distract the audience from more useful information, and encourage anxiety. We hope this case study spurs more research into a holistic analysis of public perception of health crises, and encourages collaboration between social media platforms and public health organizations.

\section{Related Works}

Advertising is the lifeblood of the majority of internet giants such as Google and Facebook -- two companies that accounted the for nearly 20\% of global advertising spending in 2016\footnote{\url{https://www.theguardian.com/media/2017/may/02/google-and-facebook-bring-in-one-fifth-of-global-ad-revenue}}. Marketing research shows that consumers have a positive attitude toward social media advertising \cite{boateng2015consumers}, though moderated by the relevance of the ad and existing perception of the company \cite{alalwan2017social}, as well as informativeness and creativity \cite{lee2016predicting}. Among the most popular social media platforms, advertising on Facebook has been shown to evoke engagement and social sharing \cite{voorveld2018engagement}. 

Having a potential reach of 2.50 billion monthly active users (MAU) as of December 2019\footnote{\url{https://zephoria.com/top-15-valuable-facebook-statistics/}}, it is not surprising that Facebook (and its other property, Instagram) has become a destination for businesses, non-profits, and politicians. Recently, the company has come under criticism for allowing political advertising of questionable quality on its platform\footnote{\url{https://techcrunch.com/2020/01/09/facebook-wont-ban-political-ads-prefers-to-keep-screwing-\\democracy}}. In response, the platform has published an Ad Library, accessible to the researchers and watchdogs to monitor advertising related to social issues. After initial reports of bugs in the API provided by Facebook to access the information in bulk \cite{rosenberg2019ad}, the library has become a valuable resource for watchdogs of political communication \cite{fischer2019facebook,masters2020were}. However, it has not yet been used in a systematic study of the relationship between such public messaging and public health attitudes. 

Beyond advertising in particular, social media in general has been a popular venue for public health campaigning, spanning efforts in smoking cessation \cite{duke2014use}, organ donor registration \cite{cameron2013social}, and sexual health promotion \cite{bull2012social}. Although it is not clear that increased online engagement results in desired health behaviors, research in commercial sphere shows a relationship between engagement with Facebook and sales \cite{brettel2015drives,kumar2013practice}. To encourage engagement, public health campaigns strive for an engaging experience, as the campaigns with a clear call to action have an opportunity to quantify the impact of the message (such as signing up for future contacts), or encourage the community to propagate the message (in best scenario having it go ``viral'', i.e. very popular) \cite{freeman2015social}. To encourage wider sharing, campaigns may personalize the messages to individuals or demographic groups \cite{noar2011tailored} or use highly engaged ``seed'' users who promote the content in their immediate social network \cite{stefanone2012click,gough2017tweet}. However, unlike in traditional advertising, sponsorships, partnerships and use of persons of authority may hurt engagement, whereas partnering with celebrities and sportspeople results in increased likes and shares \cite{kite2016please}. This trend can be attributed to a variety of reasons: perceived credibility associated with success, social acceptance, and confirmation bias, all provide positive reinforcement for people to follow health advice of celebrities, possibly propagating harmful behaviors and beliefs \cite{hoffman2015biological}.

Public perception of health issues has long been entangled with the opinions expressed by celebrities and politicians. Historically, celebrity ``health narratives'' resulted in a ``co-construction of meaning'', allowing the audience to participate in the experience of a health event and reflect on its significance in their own lives \cite{beck2015celebrity}. However, many perceptions may happen subconsciously. For example, a recent study showed that children who saw influencers with unhealthy snacks had significantly increased overall food intake, with no accompanying positive response to seeing them with healthy snacks \cite{coates2019social}. The popularization of social media and the expansion of authority and celebrity have brought informal signals about health and medicine to a vast number of internet users -- to their possible detriment. 

Early on, information available on Internet has been shown to be problematic, with a wide variability in the quality of health-related information \cite{schmidt2004assessing,dy2012effect}. The openness and scale of social media makes it especially susceptible to harmful information, such as YouTube videos promoting tobacco to consumers \cite{lau2012social}, Twitter posts sowing doubt about the safety of vaccination \cite{mitra2016understanding}, and Flickr communities ``supporting'' its members in maintaining anorexic behaviors \cite{yom2012pro}. Especially during epidemics, social media allows a rapid spread of rumors and misinformation. To track such content, a hybrid approach has been proposed wherein expert knowledge is combined with citizen science and machine learning. Such pipelines have been proposed for the 2014 Ebola \cite{fung2016social} and 2016 Zika \cite{dredze2016zika,Ghenai2017catching,daughton2019identifying} outbreaks. Resources such as HealthMap\footnote{\url{https://healthmap.org/en/}} and AIDR\footnote{\url{http://aidr.qcri.org/}} attempt to streamline the processing of news, social media, and medical reports to build an up-to-date model of an ongoing crisis \cite{kraemer2019utilizing,rudra2017classifying}. However, little has been done in understanding advertisement plays in public health communication during an epidemic. Although social media websites are attempting to apply quality control to the ads on their platforms, such as Facebook down-ranking ads mentioning ``vaccine hoaxes''\footnote{\url{https://www.theguardian.com/technology/2019/mar/07/facebook-anti-vaxx-vaccine-hoax-ads}}, until now it has been difficult to examine the quality of information posted via such paid channels. In this study, we examine the case of the 2019-2020 coronavirus epidemic via a newly available window into advertising on the Facebook platform.

\section{Data}

We begin by querying Facebook Ads Library API on February 20, 2020 with the keywords ``coronavirus'' and ''covid-19'', collecting all available ads (ongoing or finished), from all available countries. The fields returned by the API include the numerical ad identifier, span of the time in which the ad is shown, a range indicating money spent on the ad, a range of ``impressions'' the ad received, ID and name of the Facebook Page posting the ad, and the name of the funding entity of the campaign. Furthermore, the viewership is broken down into gender (3) and age (7) buckets, as well as country regions such as states in the USA. Note that the purpose of this Library is to expose ads Facebook deems relevant to ``social issues, elections or politics'', which likely do not encompass all advertising from the public health domain. Thus, we focus on the mentions of the epidemic in \emph{alternative} domains. 

\begin{figure}[t]
\centering
\includegraphics[width=0.95\columnwidth]{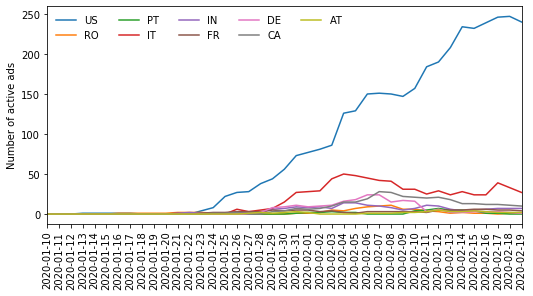}
\caption{\textbf{Number of active ads per day for countries having at least 10 ads in the dataset.} }
\label{fig:timeseriesads}
\end{figure}

In total, we collect 923 ads from 34 countries. However, only a few countries had a substantial number of ads. Figure \ref{fig:timeseriesads} shows the number of ads active per day for the countries which have have at least 10 ads in total. We find that the majority of captured advertising came from the United States, having 359 ads in our dataset, followed by Italy at 228, and India at 64. The first advertisement we find is in the United States on January 13, a news report on the emergence of a new strain of coronavirus in China. 

In this study, we focus on the advertisements originating from the United States, as this makes possible an examination of a cohesive picture of public health attitudes, as well as political context in which the ads were placed. However, we make the full dataset available to the research community, under the Terms of Service of Facebook\footnote{\url{https://developers.facebook.com/policy/}}.

\begin{figure*}[th] 
{
\centering
\includegraphics[width=0.9\linewidth]{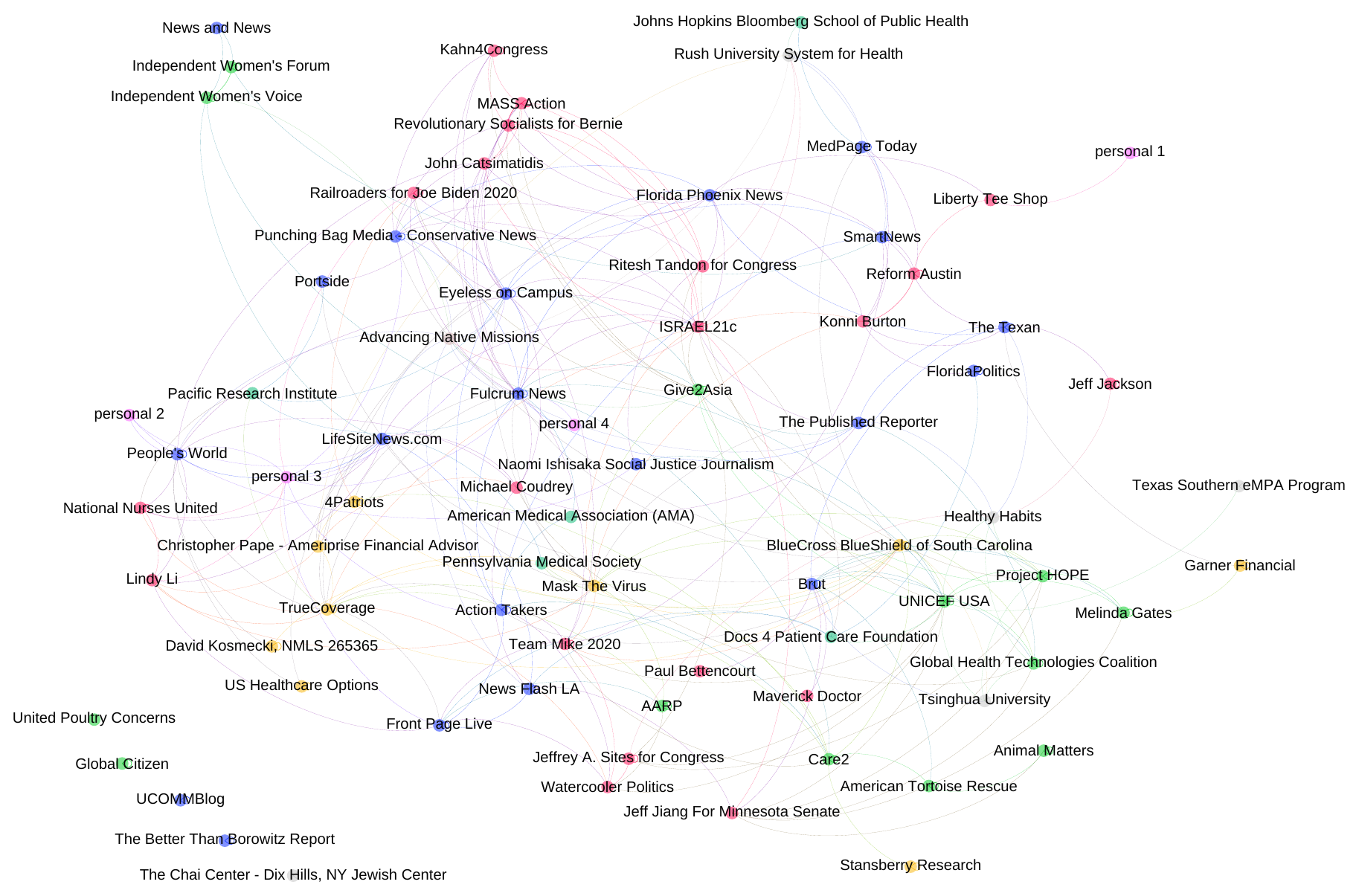}
\caption{\textbf{Network of pages, linked using Jaccard similarity of ad text at threshold 0.05. Node colors: red - politics, blue - news, light green - non-profit, dark green - public health, yellow - business, pink - personal. 
Force-directed Gephi layout Force Atlas 2.}}
\label{fig:pagenetwork}
}
\end{figure*}

\section{Results}

\subsection{Advertisers}


We begin by examining the 78 distinct Facebook Pages which have posted advertisements in the span of our dataset. We manually examine the link to their pages, as provided by the API, and note their category, as stated on the page, number of likes, and self-description, if one is provided by the page authors. We then use open coding in order to classify the pages into \emph{public health} (6\%), \emph{non-profit} (16\%), \emph{news} (26\%), \emph{political} (26\%), \emph{business} (12\%), and \emph{personal} (5\%), with the rest of categories accounting for 9\% of the advertisers. Coding result was examined by all authors and uncertain cases were decided jointly. The most difficult distinction proved to be between \emph{news} and \emph{political}, as most such pages posted politically-relevant news, thus only pages representing a political figure or party were classified as \emph{political}. Also note the small contribution of \emph{public health} to the dataset. It is possible the Library selection criteria Facebook applies to the ads simply does not match that of most public health campaigns, and a more varied resource is necessary to fully capture them. Thus, we focus on the non-public health campaigns.

Figure \ref{fig:pagenetwork} shows the pages in a co-occurrence network such that each node is a Facebook page posting an advertisement and an edge is the Jaccard similarity between the text of all of their ads in our dataset. To clean the text, we remove URLs, special characters and stopwords, and lemmatize the remaining words. For visual clarity, we threshold Jaccard similarity at 0.05. The nodes are colored by the category of the page. 
We use Gephi force-directed layout ForceAtlas 2 \cite{cherven2013network} which positions nodes which are strongly connected closer together. We notice a cluster of non-profit organizations (green nodes) in lower right, which include UNICEF, Melinda Gates Foundation, and several animal rights organizations. Note that highly specialized animal rights organization United Poultry Concerns is disconnected from this group, as they use highly specific language in reference to the treatment of birds. The rest of the network shows a mixing of news (blue) and politics (green), which is expected, given most news pages are highly political. Areas of specializations are seen in the clusters of business pages (yellow) in bottom left and candidates for political office (red) in upper left, as well as two educational institutions (green and grey) in upper right. Overall, the pages form a well-connected giant connected component, with few pages excluded, indicating that the language used in the ads is largely similar.

\begin{figure}[t]
\centering
\includegraphics[width=0.95\columnwidth]{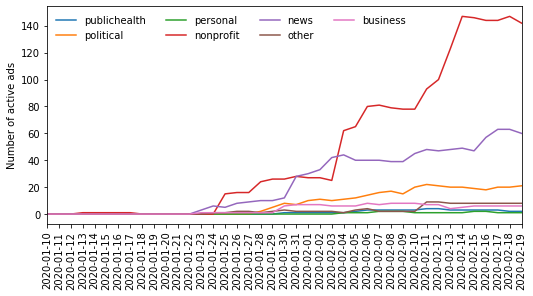}
\caption{\textbf{Number of active ads per day per page category.}}
\label{fig:ads_per_day_categories}
\end{figure}

\begin{figure*}[t]
\centering
\subfloat[Investment in USD]{\includegraphics[width=0.69\columnwidth]{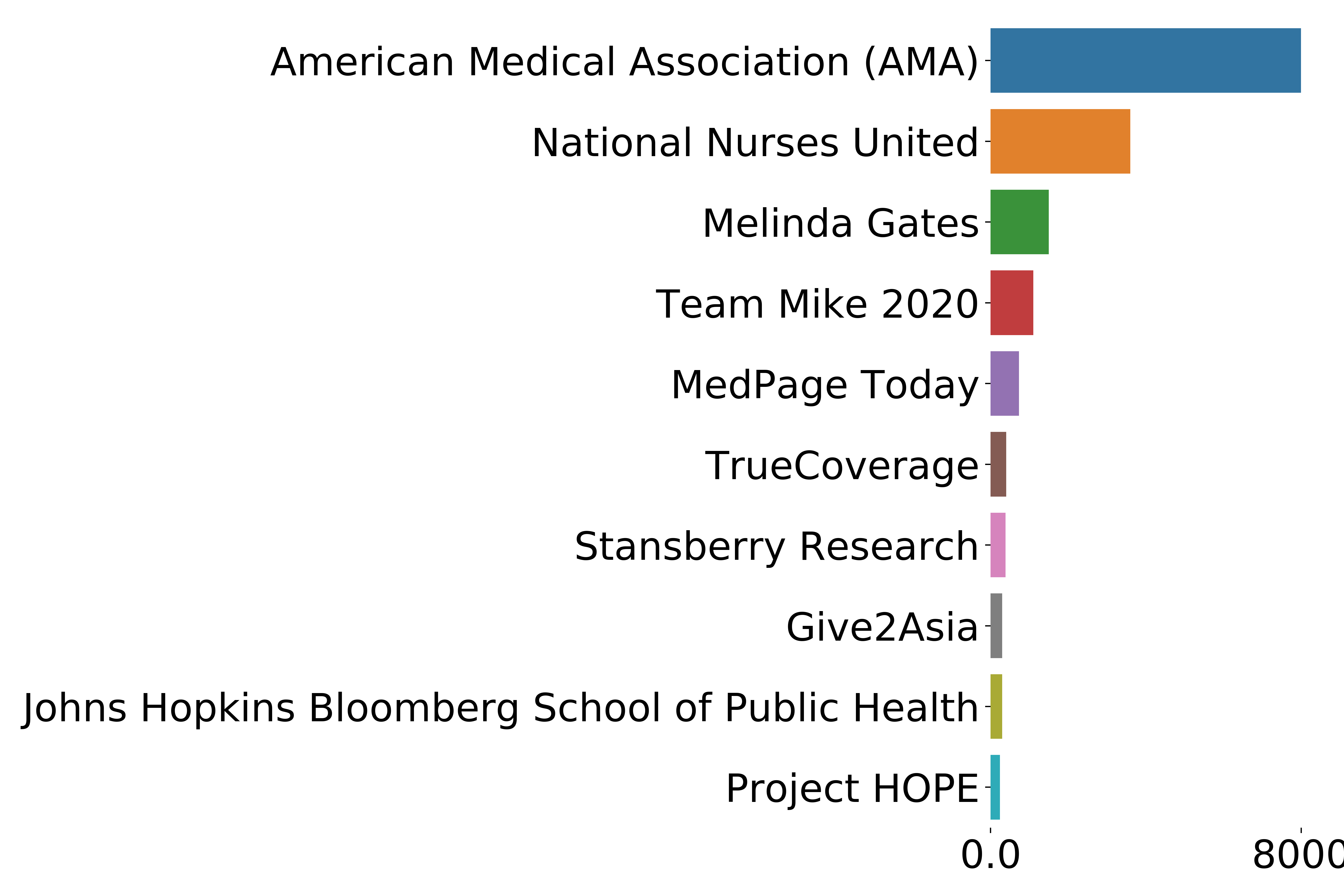}\hspace{0.15cm}}
\subfloat[Impressions]{\includegraphics[width=0.69\columnwidth]{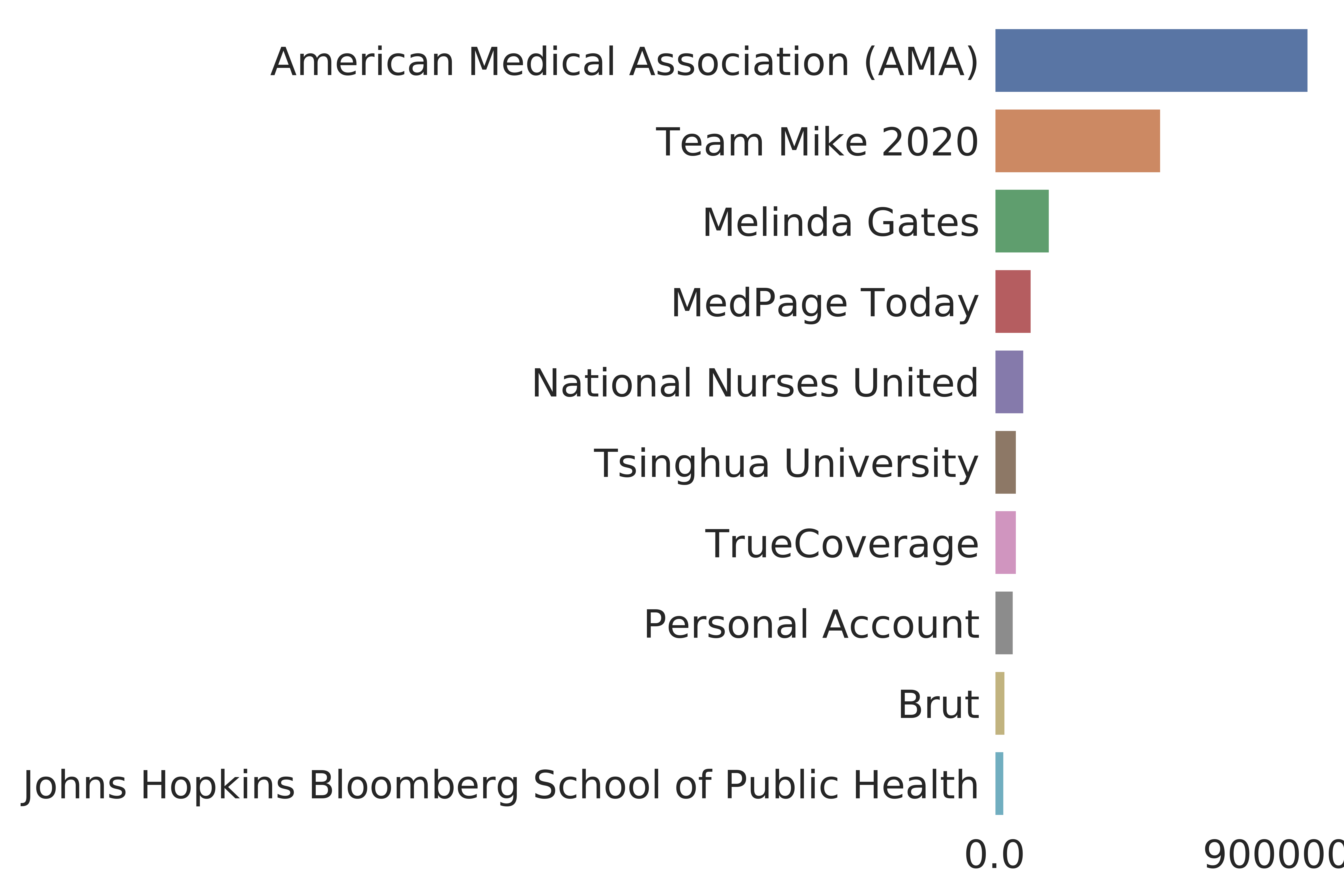}\hspace{0.15cm}}
\subfloat[Investment per Impression]{\includegraphics[width=0.69\columnwidth]{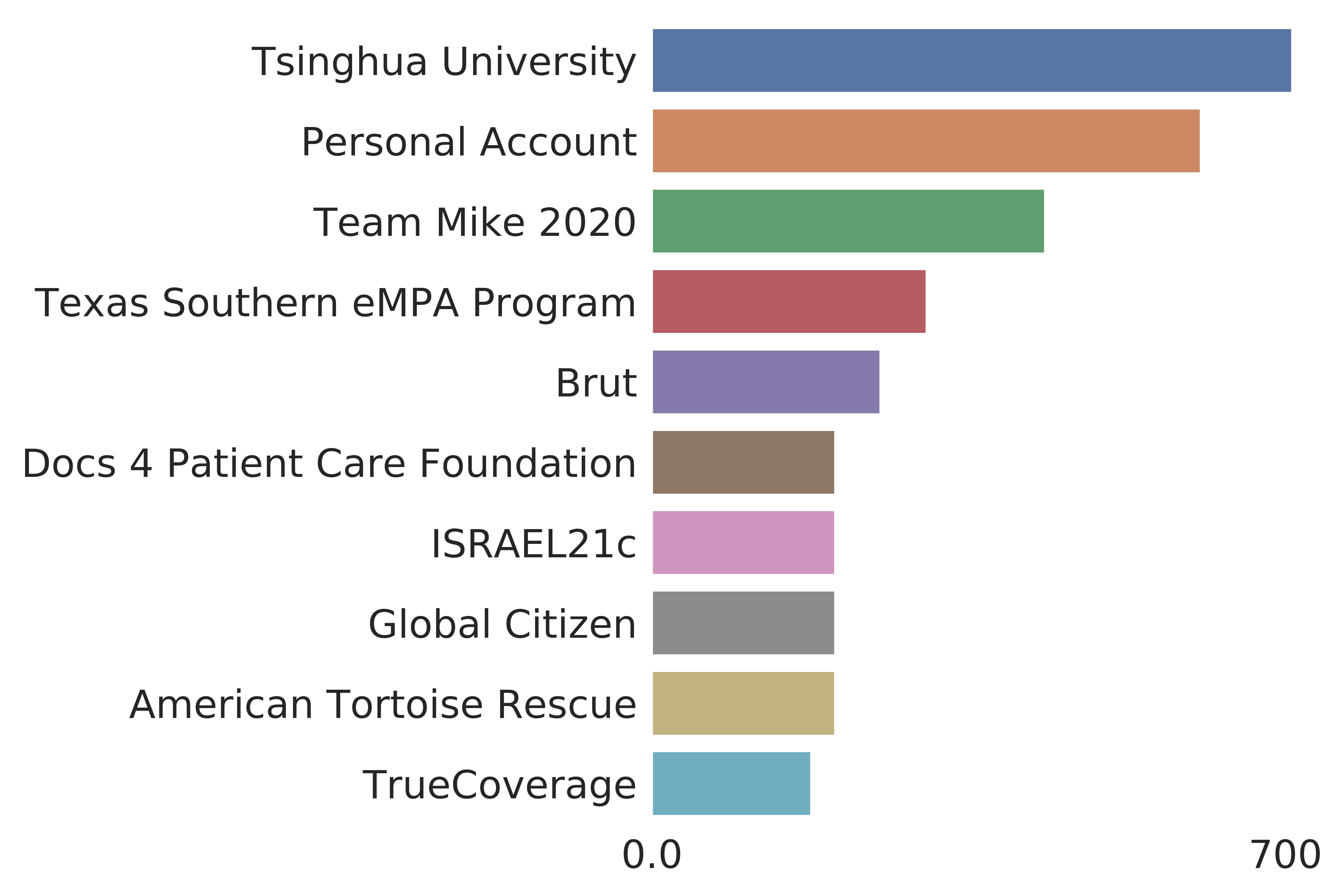}}
\caption{\textbf{Top pages by expenditures (left) and impressions (center) and dollars spent per impression (right).}}
\label{fig:pages_expend_impress}
\end{figure*}

\begin{figure*}[t]
\centering
\subfloat[Public Health]{\includegraphics[width=0.44\columnwidth]{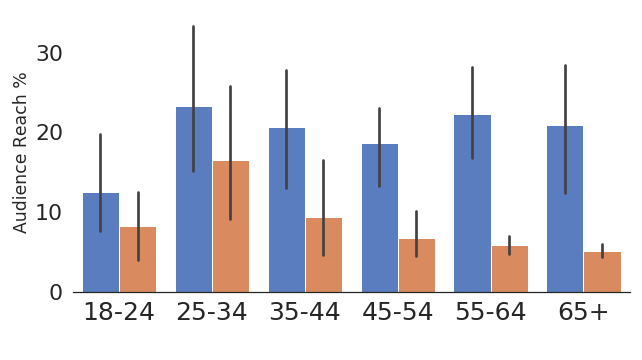}\hspace{-0.2cm}}
\subfloat[Non-Profit]{\includegraphics[width=0.44\columnwidth]{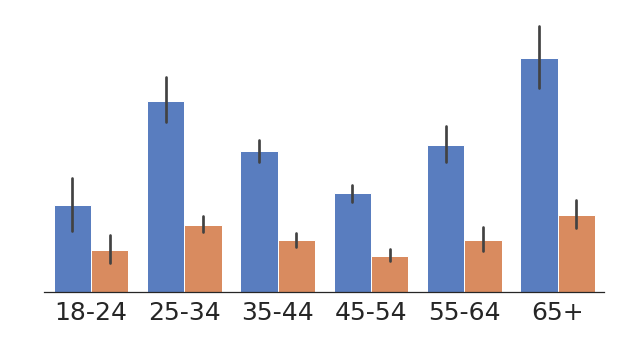}\hspace{-0.2cm}}
\subfloat[News]{\includegraphics[width=0.44\columnwidth]{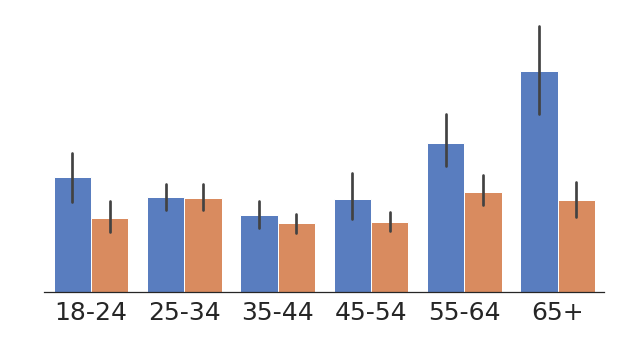}\hspace{-0.2cm}}
\subfloat[Political]{\includegraphics[width=0.44\columnwidth]{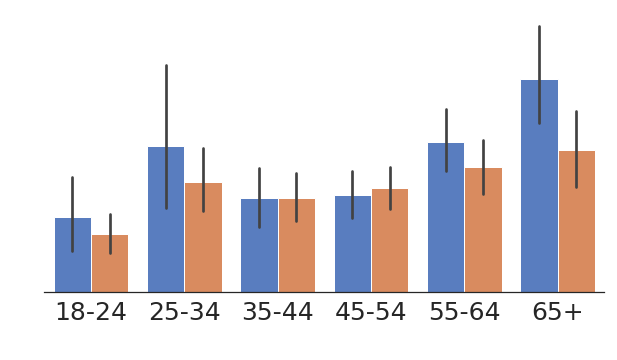}\hspace{-0.2cm}}
\subfloat[Business]{\includegraphics[width=0.44\columnwidth]{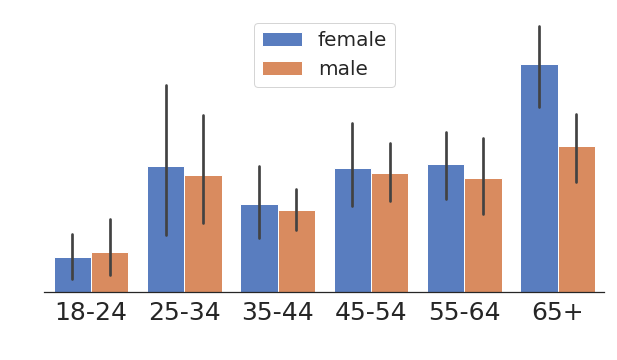}}\hspace{0.2cm}\\
\vspace{-0.1cm}
\subfloat[Public Health]{
\includegraphics[width=0.41\columnwidth]{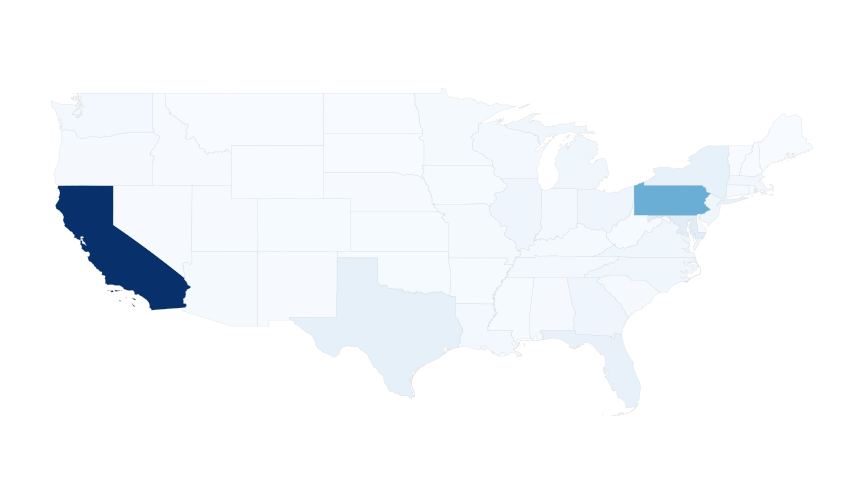}}
\subfloat[Non-Profit]{\includegraphics[width=0.41\columnwidth]{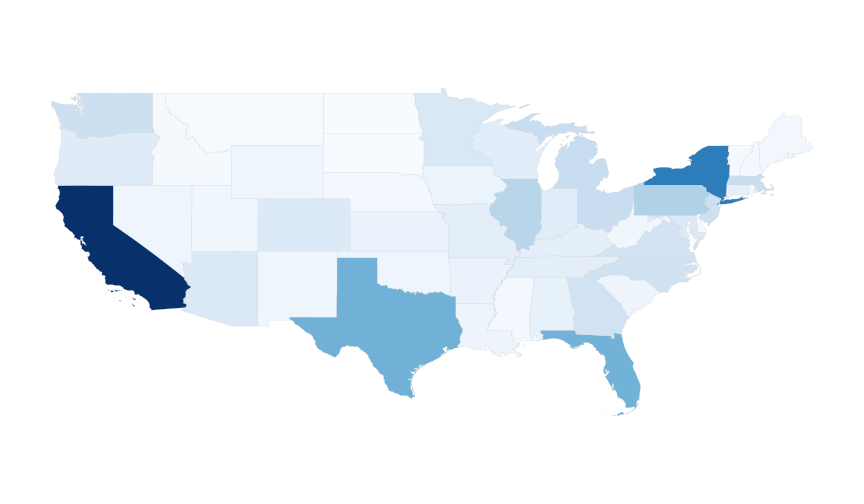}}
\subfloat[News]{\includegraphics[width=0.41\columnwidth]{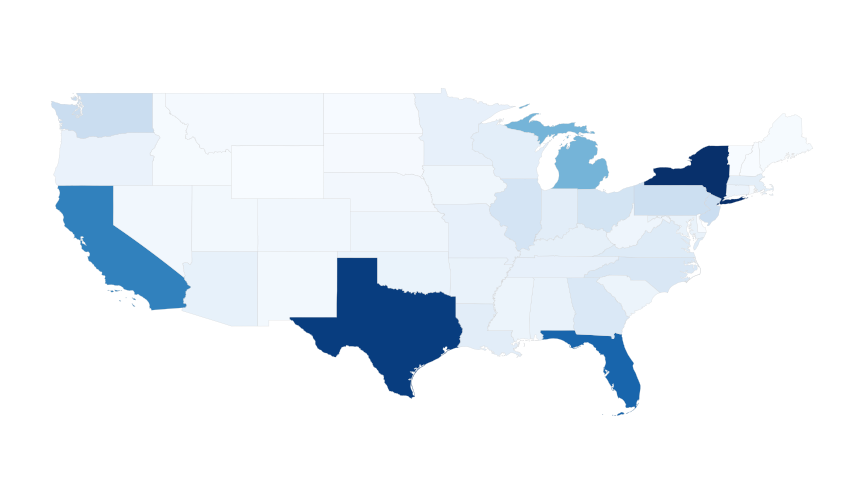}}
\subfloat[Political]{\includegraphics[width=0.41\columnwidth]{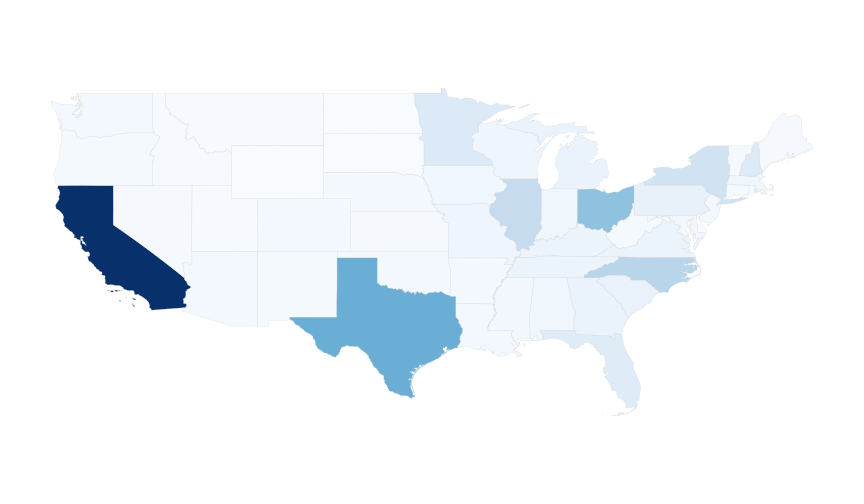}}
\subfloat[Business]{\includegraphics[width=0.41\columnwidth]{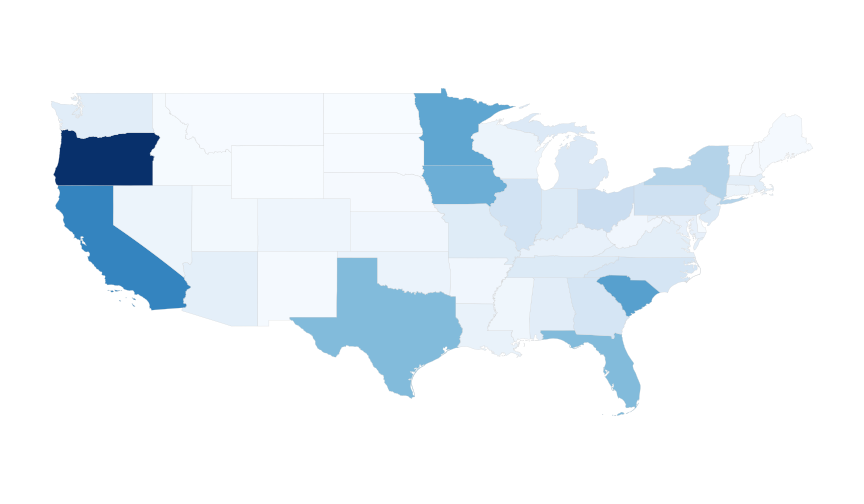}\includegraphics[width=0.08\columnwidth]{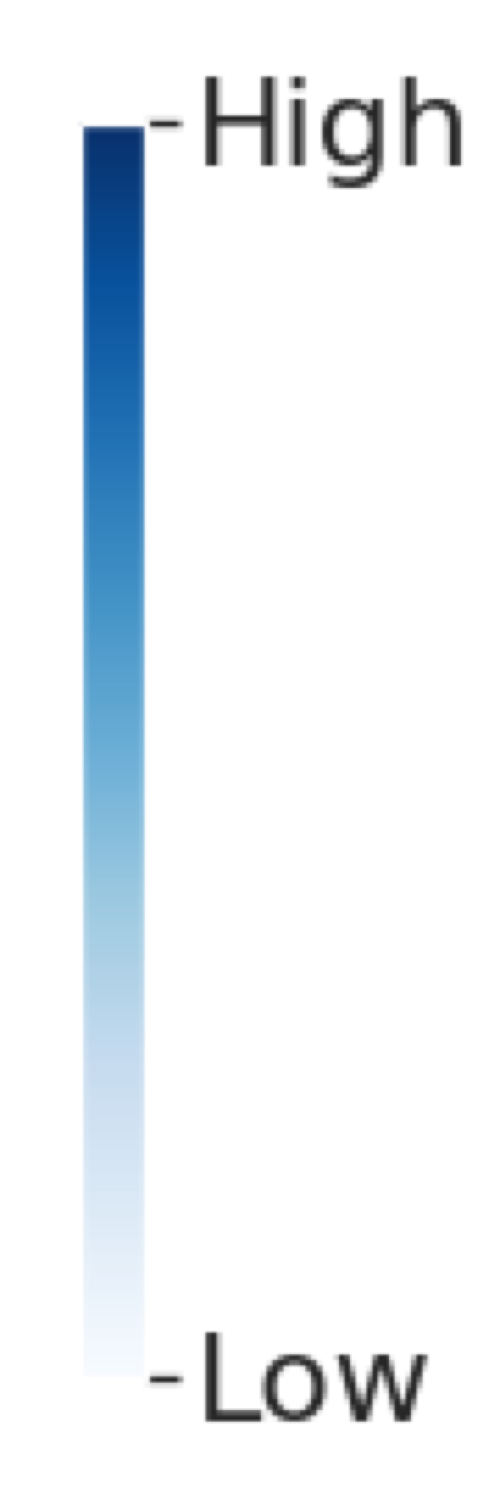}}
\caption{\textbf{Demographic and geographic targeting. From right to left: public health, non-profit, news, political, business.}}
\label{fig:demo_geo_target}
\end{figure*}

Next, we turn to the sizes of the advertising campaigns. Figure \ref{fig:ads_per_day_categories} shows the number of ads active in a day by for each page category. We find the most active to be non-profit pages, followed by news, and political ads. However, the number of impressions can vary greatly between different campaigns. The volume of non-profit pages jumps especially around February 5 when the first coronavirus case is reported in US\footnote{\url{https://kwwl.com/2020/02/05/breaking-first-case-of-\\coronavirus-confirmed-in-wisconsin/}}. In Figure \ref{fig:pages_expend_impress} we show the total expenditures (left), total impressions attained (center), and dollars per impression (center) by top pages in our dataset. As both expenditures and impressions come as ranges (such as 200-299 USD for 30,000-34,999 impressions), we take the average of min and max of each range, and sum up these averages for all ads posted by the page. Pages spending the most on their campaigns are American Medical Association (association of physicians), National Nurses United (labor union), Melinda Gates (philanthropic foundation), and Team Mike 2020 (Presidential election campaign for Michael Bloomberg). These expenditures are also associated with some of the highest audience reaches, as can be seen from the plot on the right. The advertisement with the highest expenditure is from National Nurses United, urging viewers to take a ``coronavirus preparedness survey'', asking whether their ``employers must be prepared for coronavirus''\footnote{\url{https://www.facebook.com/ads/library/?id=634538270626582}}. Yet another advertisement with high viewership and cost is by American Medical Association linking to a story of a US clinic that cared for patients infected with the coronavirus\footnote{\url{https://www.facebook.com/ads/library/?id=2456170821379373}}. We observe that most visible ads were not providing basic information about coronavirus.  Combining the two measures, in Figure \ref{fig:pages_expend_impress} (right) we show the dollar per impression spent. We observe that the highest such ratio reaches 700 USD per impression. This may be due to the actual numbers being closer to the lower end of the range than the average, which is what we employ. 

\subsection{Audience Targeting}

Along with expenditure figures for each ad, the Library provides a breakdown of the audience reached in terms of demographics (age and gender) and geography (at the state level). In attempt to reduce the effect of one advertiser posting many ads on the category statistics, we first aggregate the demographic distributions of ads for each page, and aggregate this distribution per category. The resulting demographic and the geographic targeting distributions for public health, non-profit, news, politics, and business are shown in Figure \ref{fig:demo_geo_target}. Both public health and non-profit pages reach more women then men, and the latter -- more women of age 65 and over. Public health campaigns, on the other hand, reach younger men. The gender is more balanced for business, political, and news, however all seem to reach older women more than men. It is unclear whether these distributions are attributable to the conscious targeting on the part of advertisers, or due to the peculiarities of their existing audiences. We discuss the extent to which targeting can be tracked in Discussion section.

Examining the geographical targeting visible in maps of Figure \ref{fig:demo_geo_target}, we find state-specific focus, which often falls on California, Texas, New York, Oregon, Florida, and others. Note that although some political pages concern political candidates from particular states, most other pages are geography neutral. This focus on a few states indicates purposeful geo-targeting of audiences throughout the advertising campaigns for all categories of pages. Thus we would discourage the use of this Ad Library to make country-wide assertions, and pay special attention to the targeted populations. Note that, although the ads in this dataset are supposed to be from the US, we detected several cases of audiences reached from other regions, including Scotland, England, New South Wales, and Puerto Rico.

\subsection{Narrative \& Emotion}

\begin{table}[t]
  \centering
  \begin{tabular}{p{0.95\columnwidth}}
\toprule
\textbf{Public Health} \\
podcast, public, illness, gross, science, time, care, response, treate, variety, laregly, uninforme, range, relati, info, pame, uninformed, raghavendra, pamed, honest \\\midrule
\textbf{Non-profit} \\
read, health, animal, global, wildlife, sell, vaccine, market, human, emergency, feature, healthy, highlight, breakthrough, year, turtle, pet, decade, traditional, remedy \\\midrule
\textbf{News} \\
source, newsandnews, case, base, fact, curate, risk, selection, student, spread, fibberlip, contract, add, state, proper, alternative, expertly, factual, clean, googlenews \\\midrule
\textbf{Politics} \\
cut, support, survey, campaign, local, listen, military, hack, online, economy, student, project, send, social, affect, theft, business, finally, indictment, personnel \\\midrule
\textbf{Business} \\
week, stock, market, investor, politician, lie, concern, quarter, fourth, cover, term, development, economic, sign, number, rise, impact, insurance, medical, continue \\
\bottomrule
  \end{tabular}
  \caption{\textbf{Top 20 distinguishing words for each category.}}
  \label{tab:distinguishingwords}
\end{table}

\begin{figure}[t]
\centering
\includegraphics[width=0.48\columnwidth]{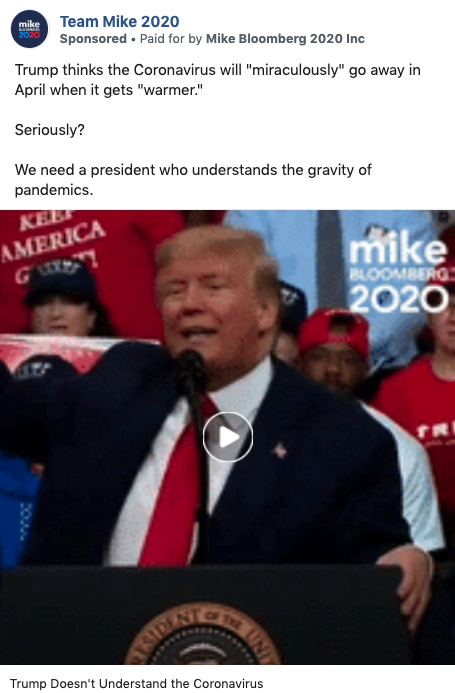}\hspace{0.2cm}
\raisebox{0.1\height}{\includegraphics[width=0.48\columnwidth]{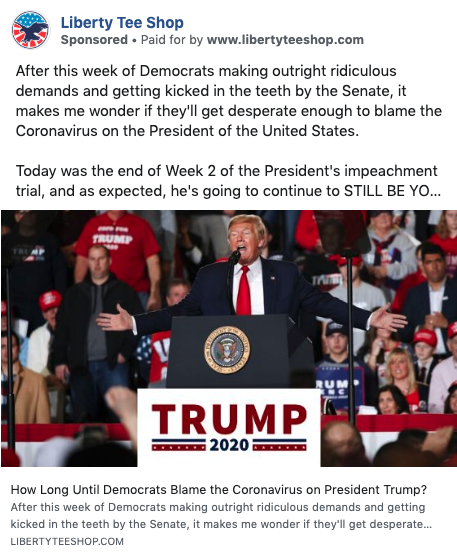}}
\caption{\textbf{Example advertisements mentioning coronavirus from political domain.} }
\label{fig:example_ad_politics}
\end{figure}

\begin{figure}[t]
\centering
\includegraphics[width=\columnwidth]{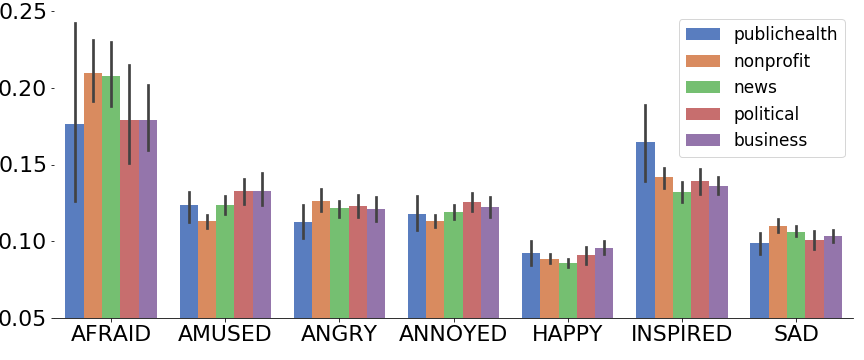}
\caption{\textbf{Mean emotions (with 95\% confidence intervals) by category.} }
\label{fig:sentiment_category}
\end{figure}

Further, we examine the language used in the ads from these categories. To do this, we consider the difference between the probability distribution (i.e. language model) of words occurring in ads of particular category and overall probability of their occurrence in the entire dataset. Table \ref{tab:distinguishingwords} lists 20 most distinguishing words for each of the major categories based on this probability difference, starting from most distinguishing ones (the score itself is not included for brevity and clarity of presentation). We observe words relevant to each of the category, such as the emphasis on science and response in public health domain, animals and wildlife in non-profit, stock market in business, support and campaigns in politics, and emphasis on particular cases and facts in news.

Despite the most funded public health-related campaign being a personal account of a doctor dealing with the disease (as discussed in the earlier section), the rest are informational articles about the currently known facts about the virus and its spread. The two main emphases of non-profit ads are donation drives and animal welfare, with some of the most seen ads coming from Melinda Gates (private charitable foundation) and Care2 (online community encouraging ``green lifestyle''). 
Most business ads come from financial advisors and insurers, who focus on the performance of the markets and whether one's insurance plan would cover coronavirus. A notable exception is a page dubbed Mask the Virus advertising masks that would ``prevent the spread of coronavirus''\footnote{\url{https://www.facebook.com/ads/library/?id=531005304180280}}. The webpage provides no description of the business, and the Facebook page was created in Jan 30, 2020 -- we discuss it at greater length in the next section. General preparedness is also promoted by a page 4Patriots (established on Jul 20, 2014), which encourages its audience to ``STOCK UP HERE''\footnote{\url{https://www.facebook.com/ads/library/?id=196433268175469}}.

The political ads come from both sides of the US political divide, such as those criticizing or supporting the US President Donald Trump (examples of such posts are in Figure \ref{fig:example_ad_politics}). As at the same time, the US is participating in primaries wherein Democratic party is choosing a contestant for US Presidential Election. Thus, ads both pro and anti several contenders also appear, mentioning Bernie Sanders, Joe Biden, Pete Buttigieg, and Michael Bloomberg, with the latter having the ad with highest impressions at over 800,000. Yet other notable campaigns come from ISRAEL21c, an organization aiming to ``inform the world about 21st century Israel'' (only such country-wide organization in our dataset), and National Nurses United, a workers union. Finally, not only US politics are mentioned: criticism of China's handling of the communications around the outbreak often takes form of the story of Li Wenliang, who was one of the first doctors to raise awareness of the new disease, and who later died of the same\footnote{\url{https://www.facebook.com/ads/library/?id=616556205810779}}.

Similar critique can be found in the ads of the news pages, however mostly the articles provide general overview of the situation, and notify public of new cases. As mentioned before, some are politically charged, and have some of the themes mentioned in political sphere. Other articles provide commentary on social impact of the situation, such as anti-asian racism and raising awareness of possible misinformation surrounding the topic. In the following section, we delve deeper into potential misinformation found in our dataset.

Lastly, we compute emotional connotation of the words used in the ads of these pages, shown in Figure \ref{fig:sentiment_category}.
We employed the DepecheMood++ lexicon
\cite{Araque2019}, which provides fine-grained emotion analysis for seven basic emotions, fear, amusement, anger, annoyance, happiness, inspiration, and sadness. As for the narrative analysis, we lemmatised the message of every advertisement and calculated the average emotion per message according to the lexicon. In this way we avoid introducing biases due to the varied lengths of the messages. 

We observe the highest levels of emotion ``afraid'', as would be expected considering the subject matter. News and non-profit ads are especially high on this emotion compared to the others. Interestingly, second most prevalent emotion is ``inspired'', with public health and non-profit showing the highest levels, possibly due to evocative messages during donation drives. As expected, ``happy'' is the least detected emotion.

\subsection{Misinformation}



During annotation of the ads, we examine the text and material to which the ad links in terms of potentially erroneous material, guided by ongoing news coverage of misinformation on mainstream news sources, fact checking sources such as Snopes and corresponding article on Wikipedia\footnote{\url{https://en.wikipedia.org/wiki/Misinformation_related_to_the_2019\%E2\%80\%9320_coronavirus_outbreak}}, as at the time of writing no comprehensive list has been provided by the public health authorities\footnote{We welcome a re-examination of ads in this dataset at a later time, when more is known about various coronavirus claims, and will make the data available for the research community.}. Out of 359 ads, we find 16 ads which have mentioned possible erroneous information, or if we consider only ads with distinct text, 8 out of 152 unique ads (5.3\%) coming mostly from categories of politics, business, and news. Also, 5 ads were debunking erroneous claims. Below, we expand on some of this content.

\begin{figure}[t]
\centering
\includegraphics[width=0.48\columnwidth]{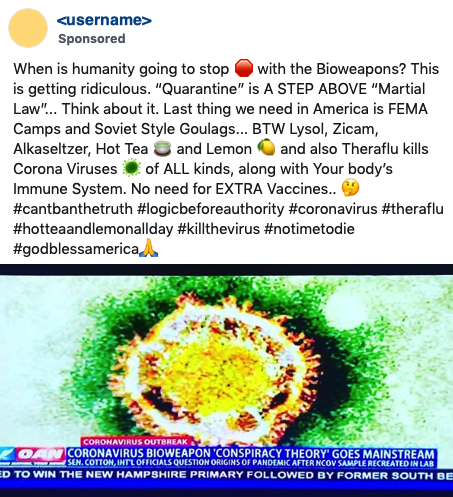}\hspace{0.2cm}
\raisebox{0.1\height}{\includegraphics[width=0.48\columnwidth]{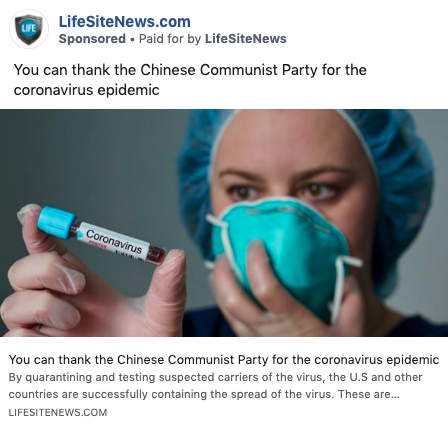}}\\\vspace{0.4cm}
\includegraphics[width=0.48\columnwidth]{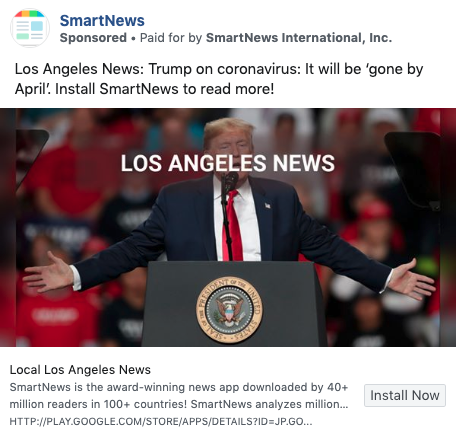}\hspace{0.2cm}
\includegraphics[width=0.48\columnwidth]{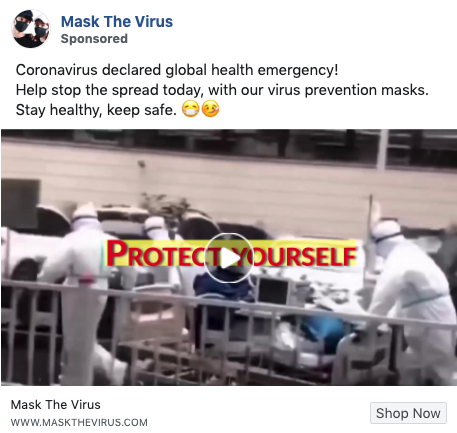}\\
\caption{\textbf{Example advertisements mentioning coronavirus and possible misinformation.} }
\label{fig:example_ad_misinfo}
\end{figure}

Perhaps the most egregious post is shown if Figure \ref{fig:example_ad_misinfo} (upper left), wherein the advertiser mentions ``bioweapons'', ``martial law'', ``FEMA camps'', ``hot tea and lemon killing corona virus'', and ``no need for extra vaccines''. As the page was created on Feb 6, 2020, it is possible that the account is a troll created specifically for posting content about this epidemic. A less obvious example is shown in Figure \ref{fig:example_ad_misinfo} (upper right), which links to an article stating ambiguously that ``there have also been suggestions that the virus may have escaped from a People’s Liberation Army Biowarfare unit'', a claim that has been circulating on Daily Mail\footnote{\url{https://www.dailymail.co.uk/health/article-7922379/Chinas-lab-studying-SARS-Ebola-Wuhan-outbreaks-center.html}} and Washington Times\footnote{\url{https://www.washingtontimes.com/news/2020/jan/26/coronavirus-link-china-biowarfare-program-possible/}}, and which the research lab is denying\footnote{\url{https://www.scmp.com/news/china/society/article/3050872/chinese-research-lab-denies-rumours-links-first-coronavirus}}. Yet less contentious, but more politicized potentially erroneous information comes from the claim by US President Donald Trump that ``as the weather starts to warm and the virus hopefully becomes weaker, and then gone'' (see Figure \ref{fig:example_ad_misinfo} (lower left)). Although the claim has some grounding in the behavior of known viruses, it is still unclear how the new strain will behave\footnote{\url{https://www.factcheck.org/2020/02/will-the-new-coronavirus-go-away-in-april/}}. Interestingly, the claim was mentioned both by news pages (simply reporting the claim), and by politically-inclined pages such as ``Team Mike 2020'', disparaging the statement. Note that beyond questionable information relating to coronavirus specifically, we detected instances of unrelated political claims and accusations captured in the political ads (though it is outside the scope of this paper to verify political statements).

\begin{figure}[t]
\centering
\subfloat[News]{\includegraphics[width=0.5\columnwidth]{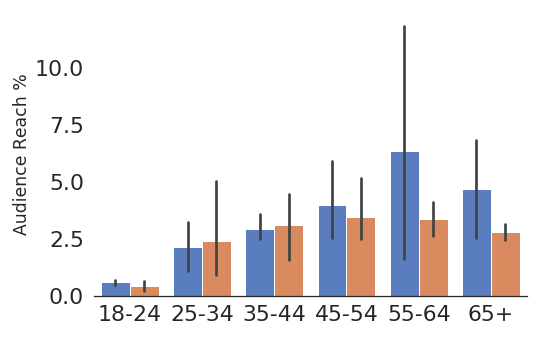}}\hspace{-0.1cm}
\subfloat[Politics]{\includegraphics[width=0.5\columnwidth]{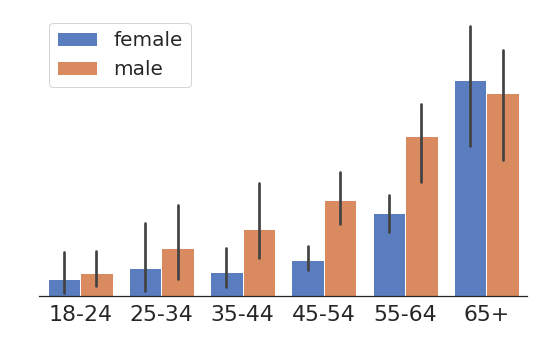}}
\caption{\textbf{Demographic breakdown of impressions of ads having possible misinformation in news and political categories.} }
\vspace{-0.4cm}
\label{fig:demog_misinfo}
\end{figure}

On the business side, we found a page created on Jan 30, 2020 to be advertising face masks (Figure \ref{fig:example_ad_misinfo} (lower right)). Neither the Facebook page or the website the ad links to provide any medical certification of the sellers of the masks, or disclose from where the business is originating\footnote{\url{https://maskthevirus.com/}}. The webpage claims these are ``Anti-Virus Masks'', and lists several versions with optional N95 and N99 activated carbon filters (which are able to filter at least 95\% of particles in the air, when mask has a tight fit). Due to limited information provided, it is difficult to determine whether these would achieve the fit necessary for protection, or whether they comply with the standards set by CDC for respirators\footnote{\url{https://www.cdc.gov/coronavirus/2019-ncov/hcp/respirator-use-faq.html}}. This is also the page having most reach outside the US. Finally, we find other topics susceptible to rumors being connected to the ongoing epidemic, such as by page ``Maverick Doctor'' linking Guillain–Barré syndrome (GBS) to vaccination, even though there is a greater risk of getting GBS after getting a flu\footnote{\url{https://www.cdc.gov/vaccinesafety/concerns/guillain-barre-syndrome.html}}. 

As can be seen from the above examples, there is a range of possible misinformation on the Facebook's advertising platform. We refer to the research community to determine whether some of these claims are intentional \emph{dis}-information, and as the crisis develops we will better understand the veracity of some of the claims. Meanwhile, we observe the myriad contexts in which the topic of coronavirus is used outside public health messaging: clear fear-mongering, political attacks, and conducting business. Although in our dataset, we find that these ads are not promoted widely by their posters, with all but one having expenditure in the range of 0-99 USD (the other being in 100-199 USD). Interestingly, the ads having possible misinformation in news and political categories have a distinct demographic targeting, as seen in Figure \ref{fig:demog_misinfo}. While news reaches age groups more evenly (except for spike for females in 55-64 age range), political ads seem to target the older generation, with highest viewership in over 65 age range. Finally, we check whether the emotional content in these advertisements is different from the rest of the dataset, and we find no statistical significance, which may indicate that misinformation indeed ``sounds'' like the rest of the content, or that we simply do not have enough data to make this assessment at this point.


\section{Discussion \& Conclusions}


At the time of data collection there were fewer than 15 cases in the United States, not counting those repatriated from other parts of the world. However, we find over 300 advertisements in a span of a month speaking about the epidemic, and potentially reaching millions of people. Many ad campaigns started even before February 5, when the first case was discovered in Wisconsin. Despite few cases, US was the most active country in the dataset. However note that many of worst affected countries in Asia are not represented in the Facebook Ads Library, so a direct comparison with them cannot be made.


Findings in this work have several implications for public health communication. Much of information provided by the news-related ads in our dataset have echoed the latest information provided by major public health organizations such as US Center for Disease Control\footnote{\url{https://www.cdc.gov/coronavirus/2019-nCoV/summary.html}}. These include information on the latest cases, travel advisories, and steps people can take to protect themselves. Some even provide links to surveys, possibly using the advertising platform as a recruitment tool for estimating awareness (although it is possible the surveys are simply meant to attract attention). Thus, Facebook advertising may be a useful way to propagate the message through alternative sources and collect data on readership. 


However, we find most advertising campaigns having extremely narrow targeting, especially geographically. Although our sample of public health campaigns is limited by Facebook's selection of ads to make available through the library, those that we find cover a small geographic locale. The situation is similar to ads from other categories. In fact, the targeting of public health and non-profit categories does not favor older men who have the highest mortality from COVID-19\footnote{\url{https://www.nytimes.com/2020/02/20/health/coronavirus-men-women.html}}. Personalized messaging has been studied in public health literature, such as for improving HIV/AIDS medication adherence \cite{dowshen2012improving} and smoking cessation \cite{naughton2013attitudes,deb2018social}, but the application of such messaging must be coordinated among the agencies to achieve a consistent message and a thorough coverage of the population. Moreover, the reach and personalization of Facebook for public health messaging can be used to reach underserved populations \cite{montague2012health,latulippe2017social}, supplementing traditional means such as physical billboards and rural outreach programs \cite{o2017improving}.


Another concern stems from the competition alternative narratives present to the messaging by public health organizations. The adage ``if it bleeds, it leads'' \cite{serani2008if} encourages news and media providers to focus on the negative and sensationalize potentially vague and uncertain knowledge about an evolving epidemic. We find that fear was the most detected emotion in ads of non-profit and news pages. Historically, epidemics such as AIDS in 1980s and 1990s \cite{gonsalves2014panic}, SARS in the early 2000s \cite{schabas2003sars}, and Ebola in 2014 \cite{parmet2017panic} tended to become associated with specific minorities and countries, with people putting the blame of the disease on the ``other''. In the case of COVID-19, the source of the virus prompted sinophobia\footnote{\url{https://www.bbc.com/news/world-asia-51456056}}, though we only find calls against racism in our dataset. 


Beside the competition for the narrative, the news, political, business and other entities are competing for the views and clicks of the audience with the public health organizations, who may not have as many resources to bid up the price. For instance, we find numerous mentions of coronavirus in association with US politicians, some of which are top spenders in our dataset. Unlike in, for example, China during SARS epidemic that prompted a ``renews sense of patriotism'' through narrative of self-sacrifice \cite{lu2008construction}, in the US COVID-19 seems to be adopted as a weapon in the ongoing partisan struggle, especially in a US Presidential election year. A similar weaponization of the epidemic narrative occurred in Africa during the Ebola epidemic, resulting in several deaths \cite{wilkinson2015briefing}. Although it is difficult to achieve a comprehensive political message during a crisis, the extent of panic amongst the public, as well as the stability of world markets, depends not only on public health institutions, but on the leaders who have far-reaching platforms.


Finally, among the advertisements in our dataset, we find about 5\% to contain possible erroneous information. Ranging from accusations of using bioweapons to questioning correctness of comparisons to other viruses, the reasons for such content may be diverse. However, we do not find the posters of these ads to be investing much money, although some achieve up to 15,000 and more impressions. It is also concerning that, based on our sentiment analysis, these messages may ``feel'' similar to the other content about coronavirus. Unfortunately, despite Facebook's collaboration with a myriad of fact-checking institutions, it is possible that small campaigns like this are not popular enough to warrant examination. The design of fully and partially automated tools to detect early appearance of ``fake news'' and misinformation is becoming a hot research topic \cite{agrawal2017multimodal,Ghenai2017catching,rajdev2015fake,vicario2019polarization}, but it is imperative to use them within a larger public health communication strategy, guided by subject matter experts, and enforced by both governments and social media platforms. 


The above conclusions must be taken with several limitations in mind. First and foremost, we are aware of the fluidity of the ongoing situation. In the light of new findings and developments, some of the annotations provided in this work may change. Thus, we will make the annotated dataset available to the research community\footnote{Please contact the first author, as Facebook Terms of Service require a signing of agreement before data can be shared. For more see \url{https://www.facebook.com/ads/library/api/}.}. Secondly, Facebook Ads Library was not created for monitoring public health messaging -- this is why this study focuses on the alternative narratives. It would be extremely helpful if the social media platforms were to collaborate more closely with public health researchers, so that a full picture of the discourse may be examined. An increased worldwide coverage would also allow the tracking of pandemics across different parts of the world (recall currently the Library covers EU countries, US, and a handful of others). However, even if we were to attain information on all advertising happening on Facebook and Instagram, this will capture by far not all internet users, and only a fraction of people in the US. Internet, as well as traditional media, present a plethora of alternative communication channels, and more studies must be conducted to capture the full scope of communication related to public health, and its possible impacts.


\textbf{Privacy.} As potentially any Facebook page may have an advertising campaign, it is possible that not only large companies and agencies, but also individuals will be captured in the Library, and in our own dataset we find several pages we labeled as ``personal''. Note that although we will make the data available in accordance with the Facebook's Terms of Service, we have chosen not to anonymize the Facebook pages other than those labeled as personal, since it is important to understand the sources of ad content, Also, all information in this dataset is freely available via Ads Library website (we even provide links to some of the ads). However, we call upon the research community to establish privacy standards for the collection, reporting, and sharing of advertising data, and we hope this study will spur the conversation in the research field. \\\vspace{2cm}

%
%
%
%
%

\balance{}

\bibliographystyle{SIGCHI-Reference-Format}
\bibliography{covid19ads}

\end{document}